\documentstyle[aps]{revtex}
\begin{document}
\draft
\title{
Phonon properties of KNbO$_3$ and KTaO$_3$ \\
from first-principles calculations
}
\author{A.~V.~Postnikov,\cite{*} T.~Neumann, and G.~Borstel}
\address{
Universit\"at Osnabr\"uck -- Fachbereich Physik,
D-49069 Osnabr\"uck, Germany}
\date{Received 1 February 1994}
\maketitle
\begin{abstract}
The frequencies of transverse-optical $\Gamma$ phonons
in KNbO$_3$ and KTaO$_3$
are calculated in the frozen-phonon scheme making use
of the full-potential linearized muffin-tin orbital
method. The calculated frequencies in the cubic phase of KNbO$_3$
and in the tetragonal ferroelectric phase
are in good agreement with experimental
data. For KTaO$_3$, the effect of lattice volume was
found to be substantial on the frequency of the soft mode,
but rather small on the relative displacement patterns
of atoms in all three modes of the $T_{1u}$ symmetry.
The TO frequencies in KTaO$_3$ are found to be of the order of,
but somehow higher than, the corresponding frequencies
in cubic KNbO$_3$.
\end{abstract}

\section{Introduction}
\label{sec:intro}
Evaluation of vibrational properties within the frozen-phonon
scheme is known to be a hard test on the quality of full-potential
total-energy calculations. The curvature of the total-energy
surface over the manifold of various atomic displacements is
much more sensitive to the details of the calculation scheme
than merely the position of the total-energy minimum, i.e.,
the equilibrium geometry. However, in case of success
(that can be easily checked by comparison with experimentally
measurable phonon frequencies) the calculation provides
substantial information on the microscopical driving forces
behind the specific vibration patterns and may give some
insight into the dynamic properties of the crystal in question.
For ferroelectric materials, the interest in the
phonon calculations is motivated by the apparently crucial role
of zone-center phonon softening in the mechanism of ferroelectric
phase transition which is subject to controversial
discussions \cite{sama87,frkc88,dwng92,sdmb93}.

An early calculation of phonon dispersion curves
within the empirical shell model
has been done for KNbO$_3$ by Fontana {\em et al.} \cite{fkc81}.
First-principles calculation of $\Gamma$ phonons
in BaTiO$_3$
which has the same crystal structure and exhibits the same
sequence of ferroelectric transitions as KNbO$_3$
have been done by Cohen and Krakauer\cite{cohen}.
Liechtenstein {\em et al.}\cite{babio3} analyzed two
vibrational modes (however not related to ferroelectric
transition) in another perovskite -- BaBiO$_3$.
Singh and Boyer\cite{sb92} calculated
$\Gamma$ and $R$ phonons in cubic KNbO$_3$.
Recently, Zhong {\em et al.}\cite{zhong}
obtained TO and LO $\Gamma$-phonon frequencies in a number
of cubic perovskite-type ferroelectrics.

In the present paper, we continue the comparative research
of KNbO$_3$ and KTaO$_3$ initiated in the previous {\it ab initio}
study of equilibrium geometry (Ref.\cite{ktn3}, referred to
further as I). In the present paper, we concentrate on
phonon frequencies in these compounds making use of
the same calculation scheme (full-potential linear muffin-tin
orbital code by Methfessel\cite{msm1,msm2}) and
setup as discussed in I.

For KNbO$_3$, we performed the $\Gamma$-phonon calculations
for the nonpolar cubic phase and for the first (as the temperature
lowers) ferroelectric phase, i.e., the tetragonal one.
The results are discussed in Secs. \ref{sec:kno_cub}
and \ref{sec:kno_tet}. In Sec. \ref{sec:kto},
the results for KTaO$_3$
in the cubic phase for two lattice spacings are presented.

\section{KN\lowercase{b}O$_3$: cubic structure}
\label{sec:kno_cub}

As has been pointed out in I, the calculated
(from the total-energy minimum) equilibrium volume of the cubic
phase of KNbO$_3$ turns out to be $\sim$95\% of the experimental
cell volume (extrapolated to zero temperature). Such discrepancy is
known to be typical for calculations based on the local density
approximation (LDA). The effect of different volumes on the trends in
total energy lowering and related equilibrium displacements is
discussed at length in I. Since the curvature
of the total-energy hypersurface may be
affected by the error of about 5\% in the cell volume,
we preferred to perform our phonon calculations for the experimental
lattice spacing, in order to produce results better
comparable with experimentally measured phonon frequencies.
Specifically, we took $a$=4.00 {\AA} for the cubic phase
(extrapolated value on $T$=0 from lattice constants of
the high-temperature phase, according to Ref.\cite{snp54}).
The measured lattice constant in the cubic phase which exists
above 418 $^0$C is about 4.02 {\AA}, slightly increasing with
temperature.\cite{snp54} The variation of volume over the
temperature range is therefore much smaller than the difference
between experimental and theoretical volumes and is ignored
in the present study.

As it is known (see, e.g., Refs.\cite{cohen} and \cite{fmsg84}),
$\Gamma$-phonon
vibration modes in the cubic perovskite structure,
after projecting out three translational modes, are split
by symmetry into three $T_{1u}$ modes and one $T_{2u}$ mode
(all of them are triple degenerate).
We followed essentially the guidelines
of Cohen and Krakauer\cite{cohen} in projecting out the translational
modes, as well as in the choice of appropriate symmetry coordinates.
When constructing the force constant matrix, we performed about 30
calculations for different displacement patterns, in order
to provide a good multidimensional fit for the total-energy
hypersurface by a fourth-order polynomial, retaining then
appropriate second derivatives.
In such a way, we could obtain stable and controllable values,
e.g., for off-diagonal elements of the force constant matrix.

The simplified procedure of extracting each of these elements
independently from a single calculation involving
particular combined displacement
has been used by Cohen and Krakauer\cite{cohen} and
applied to KNbO$_3$ by Singh and Boyer\cite{sb92}.
In our opinion, this scheme may be not sufficiently reliable
in case of an essentially nonquadratic shape
of the total-energy hypersurface, resulting in
the apparent dependence of
the calculated force constants on the particular displacement
chosen.

On solving the secular equation of the
3$\times$3 lattice dynamics problem (for the $T_{1u}$ mode),
\[
 [{\bf G}{\bf F} - \omega^2] {\bf u} = 0
\]
(Refs. \cite{wdc55} and \cite{dh77}),
with ${\bf G}$ and ${\bf F}$ matrices defined as in Ref.\cite{cohen},
the pattern of Cartesian displacements of individual atoms
is restored by back symmetry transformation.
For the atoms with the fractional coordinates
in the cubic perovskite cell
K(0, 0, 0); Nb(0.5, 0.5, 0.5);
O$_1$(0, 0.5, 0.5); O$_2$(0.5, 0, 0.5); O$_3$(0.5, 0.5, 0),
the vibrations, e.g., along the [001] direction
are given by
\[
\left(
\begin{array}{c}
  z_1 \\ z_2 \\ z_3 \\ z_4 \\ z_5
\end{array}
\right)
=
\left(
\begin{array}{rrr}
 ~4~ & -1~ & -1~ \\
 -1~ & ~4~ & -1~ \\
 -1~ & -1~ & -1~ \\
 -1~ & -1~ & -1~ \\
 -1~ & -1~ & ~4~
\end{array}
\right)
\left(
\begin{array}{c}
  u_1 \\ u_2 \\ u_3
\end{array}
\right).
\]
These displacements related to the center of mass
and multiplied by square roots of individual atomic masses
produce the orthogonal eigenvectors of vibrational modes,
which are presented, along with the frequencies,
in Table \ref{phon:knocub} in comparison with some
experimental data.

Our calculations reproduce the measured frequencies
of TO$_2$ -- TO$_4$ phonon modes reasonably well, while the TO$_1$ mode
frequency is imaginary, as can be expected for a soft mode
at zero temperature. In accordance with our previous analysis
of I and in agreement with the results of Singh and Boyer, \cite{sb92}
the eigenvector corresponding to the soft mode represents
roughly a displacement of Nb with respect to the rest of the crystal.
It should be noted that in I, the off-center displacement of Nb
was found to be energetically favourable not only at the experimental
cell volume for $a=$7.553 a.u., but also at the somewhat smaller
optimized theoretical volume ($a=$7.425 a.u.). Therefore,
the soft mode frequency, even if calculated at the latter volume,
is expected to be imaginary. On the contrary,
Singh and Boyer\cite{sb92} could only obtain the energy
lowering by Nb displacements at $a=$7.589 a.u.
(the value taken in Ref.\cite{sb92}
for the experimental lattice constant),
but not at their theoretical lattice constant of 7.448 a.u.
This discrepancy is probably related to some differences
in the calculation schemes and needs to be further
investigated.

Our results indicate a good agreement with the experimental
data in determining the frequency of the TO$_2$ mode,
whereas the frequencies of TO$_3$ and TO$_4$ modes are
systematically underestimated in both the present calculation
and that of Singh and Boyer.\cite{sb92} The reason may be
that the TO$_2$ mode is essentially the vibration of K atoms
against the rest of crystal. As soon as the potential well
related to individual off-center displacements of K is the most
parabolic comparing to potentials
felt by all other atoms (it has been discussed
at length in I), the harmonic approximation seems to work
best for this kind of displacement. Additional evidence
confirming this point is that the experimental frequencies
measured for this mode are most stable over a broad temperature
range. The reason why the corresponding frequency is
underestimated by 15\% in the calculation of Singh and Boyer\cite{sb92}
may be the unadequacy of their fit for the total-energy
hypersurface which has been constructed only on six
different displacement patterns to span all three
$T_{1u}$-type modes.

TO$_3$ and TO$_4$ modes involve two different kinds
of stretching of the oxygen sublattice which seem to
give rise to a rather unparabolic total-energy surface,
so the accuracy of the phonon description in the harmonic
approximation is not sufficient.

Turning to a more detailed analysis of the eigenvectors,
one may conclude that there is an overall agreement
between the displacement patterns calculated in the present
work and in Ref.\cite{sb92}, for the experimental lattice constant.
(There is also clear similarity with the displacement
patterns calculated for $T_{1u}$ modes of BaTiO$_3$
by Cohen and Krakauer\cite{cohen}).
However, in our calculation the displacement of K atoms
in the soft mode and in the TO$_3$ mode is more pronounced.

The displacement pattern within the soft mode,
although calculated from the second derivatives
of the total-energy hypersurface in the ideal
cubic structure, contains essential information
about relative finite displacements of atoms
in the course of cubic to tetragonal phase transition.
This has been shown already by Cohen and Krakauer\cite{cohen}
for BaTiO$_3$. For KNbO$_3$, the relative displacements
corresponding to the eigenvector of the soft mode
in our calculation are presented in Fig. \ref{fig1}(a)
in comparison with the atomic positions in the tetragonal
cell as determined by neutron diffraction\cite{hewat}
[Fig.\ref{fig1}(b)].
Whereas the experimental coordinates may be interpreted
as being due to opposite movement of Nb and O sublattices,
with the K atoms relatively undisplaced from their
positions in the center-of-mass scale, our calculations
show a pronounced tendency of K atoms to remain
stuck to the oxygen sublattice, so that the ferroelectric
transition may be roughly considered as primarily
due to net off-center Nb displacement versus the rest of the crystal.
It should be noted that the K displacement
has been experimentally estimated with the lowest accuracy,
compared to other atoms, as has been pointed out in Ref.\cite{hewat}.

\section{KN\lowercase{b}O$_3$: tetragonal structure}
\label{sec:kno_tet}
For the tetragonal phase, the situation is complicated
by the presence of strain. In I, we have shown
that the tetragonal strain can be optimized, along with
the off-center displacement of Nb atoms and under the constraint
of constant (theoretical) cell volume,
to be in fairly good agreement with
the experimental estimate of $c/a\sim$1.02. Now that we would like
to proceed with the experimental lattice constant which does not
provide the minimum of the calculated total energy, the optimization
of the strain from first principles is no longer justified.
Instead, we should take the experimental lattice constants
at some temperature as
an external constraint. The choice $a$=4.00 \AA, $c/a$=1.0165
corresponds to what is measured at about 250$^0$,
near the lowest-temperature range of existence of
the tetragonal phase.\cite{snp54}
Within thus predetermined crystal lattice,
associated with any one type of atom (for instance, Nb),
K and two nonequivalent types of O are free to relax
along the [001] axis towards equilibrium positions
compatible with tetragonal symmetry. Since this equilibrium
geometry is not determined by symmetry as in the cubic phase,
it must be found from first-principles calculations prior to
further frozen-phonon analysis. This search for the global
total-energy minimum over three independent parameters has been
accomplished by a polynomial fitting and needed several tens
of total-energy calculations to achieve sufficient accuracy.

The optimized perturbations of the fractional coordinates
of atoms in the cubic perovskite cell, accounting for
the off-center displacements along [001] in the tetragonal structure,
are found to be the following:
0.046(K); 0(Nb); 0.034(O$_1$, O$_2$); 0.047(O$_3$).
The arbitrary choice of unshifted Nb sublattice
is taken here in order to enable a direct comparison with
the experimental data of Hewat\cite{hewat} which are
correspondingly 0.018; 0; 0.040; and 0.044.
The absolute displacements of atoms from their
symmetry positions are shown in Fig. \ref{fig1}(c),
where the center of mass has been kept fixed.

As was the case with the soft mode analysis in Sec.
\ref{sec:kno_cub}, K shows, according to our calculation,
a pronounced tendency to remain stuck to the
oxygen sublattice. This is identical to what
we have found in I for the rhombohedral phase,
when considering the energetics of coupled
K and O distortions.
Otherwise, the main trend in the ferroelectric transition,
namely that the largest displacement is that of Nb with respect
to O$_3$, as well as the magnitude of this displacement,
are in fairly good agreement with the experiment.

As the crystal space group is reduced from
 $Pm3m$  to $P4mm$
at the ferroelectric phase transition, the vibrations along
the [001] direction are no more degenerate; three corresponding
$T_{1u}$ modes which retain the tetragonal symmetry of the crystal lattice
belong now to the $A_1$ representation, whereas the formerly
$T_{2u}$ mode now becomes $B_1$. After projecting out
the uniform displacement along [001], we arrive at the following
symmetry coordinates (similar to those of Ref.\cite{cohen}):
\[
\begin{array}{c}
 \begin{array}{ll}
   A_1 & \left\{
    \begin{array}{c}
     ~\\~\\~
    \end{array}
    \right.
\end{array} \\
\text{transl.} \\
 \begin{array}{ll}
  B_1 & \left\{ ~ \right.
 \end{array}
\end{array}
\left(
\begin{array}{l}
  S_1 \\ S_2 \\ S_3 \\ ~ \\ S_4
\end{array}
\right)
=
\left(
\begin{array}{rrrrr}
 ~1~~ & ~0~~ & ~0~~ & ~0~~ & -1~ \\
 ~0~~ & ~1~~ & ~0~~ & ~0~~ & -1~ \\
 ~0~~ & ~0~~ & ~0.5 & ~0.5 & -1~ \\
 ~1~~ & ~1~~ & ~1~~ & ~1~~ & ~1~ \\
 ~0~~ & ~0~~ & ~1~~ & -1~~ & ~0~
\end{array}
\right)
\left(
\begin{array}{l}
  z_1 \\ z_2 \\ z_3 \\ z_4 \\ z_5
\end{array}
\right).
\]

Vibrations normal to the tetragonal axis remain doubly
degenerate; former $T_{1u}$ and $T_{2u}$ modes however
are now mixed in the $E$ representation, with the following
symmetry coordinates (along [100]):
\[
\begin{array}{c}
 \begin{array}{ll}
   E & \left\{
    \begin{array}{c}
     ~\\~\\~\\~
    \end{array}
    \right.
 \end{array} \\
\text{transl.}
\end{array}
\left(
\begin{array}{l}
  S_1 \\ S_2 \\ S_3 \\ S_4 \\ ~
\end{array}
\right)
=
\left(
\begin{array}{rrrrr}
 ~1~ & ~0~ & ~0~ & ~0~ & -1~ \\
 ~0~ & ~1~ & ~0~ & ~0~ & -1~ \\
 ~0~ & ~0~ & ~1~ & ~0~ & -1~ \\
 ~0~ & ~0~ & ~0~ & ~1~ & -1~ \\
 ~1~ & ~1~ & ~1~ & ~1~ & ~1~
\end{array}
\right)
\left(
\begin{array}{l}
  x_1 \\ x_2 \\ x_3 \\ x_4 \\ x_5
\end{array}
\right)
\]
and analogously along [010].

Calculated phonon frequencies and restored orthogonal eigenvectors
(individual atomic displacements,
multiplied by square roots of masses) are presented
in Table \ref{calc:knotet}; a comparison with the experimentally
determined frequencies is shown in Table \ref{exp:knotet}.
With the exception of the soft TO$_1$ mode, experimental
frequencies are fairly stable within the temperature range
of the tetragonal phase, and calculated frequencies are
in all cases in reasonable agreement with them. The best agreement
is obtained for the modes involving a relatively
small amount of Nb vibration, which is known from our
previous analysis in I to be essentially anharmonic. The most parabolic
potential well, as has been already emphasized
in Sec. \ref{sec:kno_cub},
is that related to the K atom; as a result,
the harmonic frequency of the TO$_2$-$E$
mode, which represents roughly the vibrations of K against
the rest of crystal, is in fairly good agreement with
experiment. The same applies to the TO$_3-A_1$ mode
which is essentially the vibration of the basal O atom
against all others. Consistent with this point of view,
the experimental frequencies for these particular modes
are the most stable over the temperature.
Other modes either include
a considerable contribution of the Nb displacement
(two lowest $A_1$ modes), or represent essentially
the stretching of the oxygen cage ($B_1$ and two highest $E$ modes),
in which case the harmonic approximation is less accurate,
and the temperature dependence of frequencies should be
noticeable.

The double degenerate soft mode exhibits the tendency
of atoms to shift from their positions,
which were optimized with respect to displacements
along [001], but correspond to
a saddle point of more general total-energy
hypersurface (see I)
in one or another orthogonal direction,
compatible with orthorombic crystal structure.
The eigenvector of the soft mode indicates that
such transformation involves primarily the displacement
of Nb sublattice against the rest of crystal, as was
the case in the cubic to tetragonal phase transition.

\section{KT\lowercase{a}O$_3$}
\label{sec:kto}

KTaO$_3$, in contrast to KNbO$_3$,
does not undergo a ferroelectric phase
transition, remaining
in the cubic perovskite structure
over the whole temperature range.
This crystal seems to be however at the
very threshold to a ferroelectric phase transition,
as is indicated by considerable softening of its TO$_1$ mode
at low temperatures (see, e.g.,
Refs.\cite{snm67} and \cite{ahs70}) and
by the fact that this transition can be induced by
applied uniaxial stress.\cite{us75}
In I, we studied the energetics of off-center displacements of
different atoms in KTaO$_3$ and found the corresponding
potential well to be the most anharmonic and volume-dependent
for Ta. We found the cubic phase to be stable at the
theoretical (i.e., underestimated by
$\sim$5\% as is typical with the LDA) volume,
but unstable towards off-center Ta displacements
at the experimental volume. In order to study in more
detail the effect
of combined atomic displacements near the ferroelectric threshold,
we calculated the phonon
frequencies and eigenvectors at both theoretical
($a$=3.928 \AA) and experimental ($a$=3.983 \AA) values
of the lattice constant.
The calculation setup
(basis set, choice of radii) was the same as described in I;
the symmetry analysis for phonons is identical to that
in the cubic phase of KNbO$_3$.

The calculated results are shown in Table \ref{phon:ktocub}
in comparison with the frequencies measured by
several techniques. Some more experimental data,
which however fall within the same limits, may be found
in Refs.\cite{snm67} and \cite{pt69}.
The above-mentioned tendency of Ta to go off center,
as found in the calculations performed for the experimental lattice
constant, results in imaginary frequency of the soft mode,
the displacement pattern of which is essentially the movement
of Ta against the rest of crystal. At the theoretical volume,
which is smaller, the soft mode frequency was found to be real,
numerically close to the data
experimentally measured at about 200 K,
and consequently higher than the experimental results
obtained at lowest temperatures. The real behavior
of the soft mode seems to fall within the limits provided
by these two cases.

In spite of the differences in the calculated soft mode
frequency, the corresponding relative displacements
of atoms determined for
both lattice constants are very close.
One can note however the difference in the vibration of
the oxygen sublattice. At the theoretical lattice constant,
i.e., for the case of a nonpolar phase and real frequency,
the oxygen cage remains rigid on the vibrations; as the volume increases
and the ferroelectric transition develops, O$_3$ atoms
shift further towards Ta ions than O$_{1,2}$ atoms do.
The resulting slight distortion of the oxygen cage is typical for
ferroelectric KNbO$_3$ according to our calculation
(Table \ref{phon:knocub}) and for BaTiO$_3$.\cite{cohen}
Similar trends are seen in the calculation by
Singh and Boyer \cite{sb92} for two lattice constants
of KNbO$_3$.

The volume dependence of the frequencies
of two higher $T_{1u}$ modes, although much less dramatic
than it was for the soft mode,
is nevertheless noticeable. The reason for the fact that
frequencies at larger volume are systematically lower
is that the potential wells related to off-center displacements
of all constituent atoms become broader as volume increases,
with smaller curvature at the equilibrium, as has been studied in I.

We cannot provide an explanation for the opposite tendency
found for the $T_{2u}$ mode. It seems not to be an artifact
of our calculation because Singh and Boyer \cite{sb92}
obtained the same trends in their calculations for two
lattice constants in KNbO$_3$.

The whole set of calculated frequencies
lies slightly but systematically higher than those in KNbO$_3$.
KTaO$_3$ therefore can be regarded
to be the stiffer crystal
as is also evident from its higher bulk modulus
(calculated in I).

\section{Summary}

We calculated frequencies and eigenvectors of TO $\Gamma$ phonons
in cubic and tetragonal KNbO$_3$ and in cubic KTaO$_3$.
For cubic KNbO$_3$, the soft $T_{1u}$ mode
is characterized by an atomic
displacement pattern which is close to the pattern of
the off-center displacements in the ferroelectric tetragonal phase,
as determined by explicit structure optimization. The frequencies
of the other two $T_{1u}$ modes and the $T_{2u}$ mode are in good
agreement with experimental data. In the tetragonal phase,
the softening of an $E$ mode indicates an instability towards
the transition to the orthorombic ferroelectric phase.
The frequencies of the other six $A_1$, $B_1$, and $E$ modes are found
from the calculation to be in reasonable agreement with
experiment -- especially for modes
whose atomic displacements mostly lie in quasi-harmonic regions
of the energy hyperspace.
In KTaO$_3$, the TO frequencies and eigenvectors have been calculated
for two values of the lattice constant in the cubic phase.
Although the tendency for the ferroelectric transition
and correspondingly the calculated frequency of the soft
mode are found to be quite sensitive to the volume,
the displacement patterns within the soft mode and
other $T_{1u}$ and $T_{2u}$ modes are not.
The frequencies are found to be close to, but
systematically slightly higher than, the corresponding
frequencies in KNbO$_3$.

\acknowledgements

The authors are grateful to M. Methfessel for his assistance and
advise in using his full-potential LMTO code, and to
A. Liechtenstein and R. Cohen for the consultations
related to frozen-phonon calculations.
Financial support of the Deutsche Forschungsgemeinschaft
(SFB~225, Graduate College) is gratefully acknowledged.

\begin{figure}
\caption{
Relative displacements of atoms in KNbO$_3$ along [001]:
from the soft mode eigenvector in the cubic structure
[(a), arbitrary scale];
from neutron diffraction measurements in the
tetragonal structure, Ref. \protect\cite{hewat} (b);
from geometry optimization in the tetragonal structure (c).
}
\label{fig1}
\end{figure}

\begin{table}
\caption{
Calculated $\Gamma$-TO frequencies and eigenvectors in cubic KNbO$_3$.
}
\label{phon:knocub}
\begin{tabular}{ccdddddcccccc}
 & &
\multicolumn{5}{c}{Eigenvectors (present work)}&
\multicolumn{3}{c}{$\omega$ calc. (cm$^{-1}$)} \\
\cline{3-7} \cline{8-10}
\raisebox{2.5ex}[0pt]{Mode}  &
\raisebox{2.5ex}[0pt]{symm.} &
  K & Nb & O$_1$ & O$_2$ & O$_3$
 &Present&Ref.\cite{sb92}$^a$&Ref.\cite{zhong}&
\multicolumn{3}{c}{
 \raisebox{2.5ex}[0pt]{$\omega$ exp. (cm$^{-1}$)}
                  } \\
\hline
TO$_1$ & $T_{1u}$ &   0.32&$-$0.67&0.29&   0.29&   0.53&
  203$i$ &115$i$&143$i$  & 96$^b$&115$^c$&139$^d$   \\
TO$_2$ & $T_{1u}$ &$-$0.81&   0.12&0.36&   0.36&   0.27&
  193    &168   &188     &198$^b$&207$^c$&203.5$^d$ \\
TO$_3$ & $T_{1u}$ &   0.13&$-$0.14&0.45&   0.45&$-$0.75&
  459    &483   &506     &521$^b$&522$^c$&511$^d$   \\
TO$_4$ & $T_{2u}$ &   0.  &   0.  &1.  &$-$1.  &   0.  &
  234    &266   &        &       &280$^e$ \\
\end{tabular}
$^a$ for $a$=4.016 \AA.\\
$^b$ Infrared reflectivity measurements
     at  710 K, Ref. \cite{fmsg84}. \\
$^c$ Infrared reflectivity measurements
     at  740 K, Ref. \cite{glsq80}. \\
$^d$ Infrared reflectivity measurements
     at 1180 K, Ref. \cite{fmsg84}. \\
$^e$ Infrared reflectivity measurements
     at  585 K (in the tetragonal phase), Ref. \cite{fmsg84}.
\end{table}

\begin{table}
\caption{
Calculated $\Gamma$-TO frequencies and eigenvectors in tetragonal KNbO$_3$.
}
\label{calc:knotet}
\begin{tabular}{ldddddd}
 & \multicolumn{1}{c}{Frequency} & \multicolumn{5}{c}{Eigenvectors} \\
\cline{3-7}
 \raisebox{2.5ex}[0pt]{Symm.} & (cm$^{-1}$) &
 K & Nb & O$_1$ & O$_2$ & O$_3$ \\
\hline
 &  167    & $-$0.88 &    0.45 &    0.12 &    0.12 &    0.05 \\
 $A_1$
 &  330    & $-$0.08 & $-$0.49 &    0.61 &    0.61 &    0.09 \\
 &  607    & $-$0.09 & $-$0.20 & $-$0.16 & $-$0.16 &    0.95 \\
\hline
 $B_1$
 &  248    &    0.   &    0.   &    1.   & $-$1.   &    0.   \\
\hline
 &  166$i$ & $-$0.35 &    0.67 & $-$0.52 & $-$0.34 & $-$0.21 \\
 &  188    & $-$0.80 &    0.10 &    0.28 &    0.40 &    0.34 \\
 \raisebox{2.5ex}[0pt]{$E$ $\parallel [100]$}
 &  234    & $-$0.00 &    0.06 & $-$0.02 &    0.64 & $-$0.76 \\
 &  456    &    0.13 & $-$0.14 & $-$0.75 &    0.48 &    0.41 \\
\end{tabular}
\end{table}

\begin{table}
\caption{
Calculated and measured frequencies of $\Gamma$-TO phonons
in tetragonal KNbO$_3$.
}
\label{exp:knotet}
\begin{tabular}{cccccc}
 & & Calculated \\
\raisebox{2.5ex}[0pt]{Mode}  &
\raisebox{2.5ex}[0pt]{Symm.} &
 frequency (cm$^{-1}$) &
\raisebox{2.5ex}[0pt]{Expt.$^a$}  &
\raisebox{2.5ex}[0pt]{Expt.$^b$}  &
\raisebox{2.5ex}[0pt]{Expt.$^c$}  \\
\hline
  & $E$ & 166$i$ &  78 &  53 \\
 \raisebox{2.5ex}[0pt]{TO$_1$} &
  $A_1$ &    330 & 295 & 275 & 280 \\
  & $E$ & 188 & 199 & 193 & 190 \\
 \raisebox{2.5ex}[0pt]{TO$_2$} &
  $A_1$ & 167 & 190 &     & 200 \\
  & $E$ & 456 & 518 &     & 590 \\
 \raisebox{2.5ex}[0pt]{TO$_3$} &
  $A_1$ & 607 & 600 &     & 600 \\
  & $E$ & 234 & 280 &     & 285 \\
 \raisebox{2.5ex}[0pt]{TO$_4$} &
  $B_1$ & 248 &     &     & 290 \\
\end{tabular}
$^a$ Infrared spectroscopy at 585 K; Ref. \cite{fmsg84}. \\
$^b$ Neutron  spectroscopy at 518 K (Ref. \cite{fdkc79});
     numerical values cited in Ref. \cite{fkmc81}. \\
$^c$ Raman    spectroscopy at 543 K; Ref. \cite{fkmc81}.
\end{table}

\begin{table}
\caption{
TO frequencies and eigenvectors in KTaO$_3$.
}
\label{phon:ktocub}
\begin{tabular}{cdddddcccc}
 &
\multicolumn{5}{c}{Eigenvectors (calculated)} &
\multicolumn{4}{c}{$\omega$ (cm$^{-1}$) } \\
\cline{2-6}
\cline{7-10}
\raisebox{2.5ex}[0pt]{Symm.}  &
  K & Ta & O$_1$ & O$_2$ & O$_3$ &
Calc. & Expt.$^a$ & Expt.$^b$ & Expt.$^c$ \\
\hline
\multicolumn{10}{c}{$a$=3.928 \AA} \\
$T_{1u}$ &    0.68 & $-$0.56 & 0.28 &    0.28 &    0.27 &
  ~71~ \\
$T_{1u}$ & $-$0.62 & $-$0.10 & 0.51 &    0.51 &    0.29 &
  227~ \\
$T_{1u}$ &    0.12 &    0.02 & 0.32 &    0.32 & $-$0.89 &
  566~ \\
$T_{2u}$ &    0.   &    0.   & 1.   & $-$1.   &    0.   &
  294~ \\
\hline
\multicolumn{10}{c}{$a$=3.983 \AA} \\
$T_{1u}$ &     0.57  & $-$0.57 & 0.31 &    0.31 &    0.40 &
  ~61$i$ & 25$-$106& ~81 & ~85 \\
$T_{1u}$ &  $-$0.72  & $-$0.02 & 0.45 &    0.45 &    0.29 &
  205~   &196$-$199& 199 & 198 \\
$T_{1u}$ &     0.14  & $-$0.04 & 0.38 &    0.38 & $-$0.83 &
  504~   &551$-$550& 546 & 556 \\
$T_{2u}$ &     0.    &    0.   & 1.   & $-$1.   &    0.   &
  330~   &         & 279 \\
\end{tabular}
$^a$ Infrared reflectivity measurements at 12$-$463 K,
    Ref. \cite{pmn67}. \\
$^b$ Hyper-Raman scattering measurements at room temperature,
    Ref. \cite{vu84}. \\
$^c$ Raman scattering measurements at room temperature (soft mode)
    and at 10 K, Ref. \cite{fw68}.
\end{table}

\setlength{\unitlength}{0.5cm}
\begin{picture}(30,40)
\put(0,0){
 \begin{picture}(30,40)
 \put(15,28){                  
   \begin{picture}(0,0)
   \put(-11, 6  ){\makebox(1,1)[c]{\Huge a }}
   \put(-13,0){\line(1,0){26.0}}
   \put(  4.82,0.0){\line(0,1){1.0}}
   \put(  3.82,1.0){\makebox(2,2)[c]{\Large\sf K }}
   \put( 12.46,0.0){\line(0,1){4.0}}
   \put( 11.46,4.0){\makebox(2,2)[c]{\Large\sf ~O$_3$ }}
   \put(  6.82,0.0){\line(0,1){4.0}}
   \put(  5.82,4.0){\makebox(2,2)[c]{\Large\sf ~O$_{1,2}$ }}
   \put( -6.53,0.0){\line(0,1){3.0}}
   \put( -7.53,3.0){\makebox(2,2)[c]{\Large\sf Nb }}
   \end{picture}
 }
 \put(15,17){                  
   \begin{picture}(0,0)
   \put(-11, 6  ){\makebox(1,1)[c]{\Huge b }}
   \put(-13,0){\line(1,0){26.0}}
   \put(-10,-0.5){\line(0,1){0.5}}
   \put(-12,-2.5){\makebox(4,2)[c]{\large $-$0.1\AA }}
   \put( 10,-0.5){\line(0,1){0.5}}
   \put(  8,-2.5){\makebox(4,2)[c]{\large    0.1\AA }}
   \put(  1.40,0.0){\line(0,1){1.0}}
   \put(  0.40,1.0){\makebox(2,2)[c]{\Large\sf K }}
   \put( 11.70,0.0){\line(0,1){4.0}}
   \put( 10.70,4.0){\makebox(2,2)[c]{\Large\sf ~O$_3$ }}
   \put(  9.90,0.0){\line(0,1){4.0}}
   \put(  8.90,4.0){\makebox(2,2)[c]{\Large\sf ~O$_{1,2}$ }}
   \put( -6.30,0.0){\line(0,1){3.0}}
   \put( -7.30,3.0){\makebox(2,2)[c]{\Large\sf Nb }}
   \end{picture}
 }
 \put(15,5){                  
   \begin{picture}(0,0)
   \put(-11, 6  ){\makebox(1,1)[c]{\Huge c }}
   \put(-13, 0  ){\line(1,0){26.0}}
   \put(-10,-0.5){\line(0,1){0.5}}
   \put(-12,-2.5){\makebox(4,2)[c]{\large $-$0.1\AA }}
   \put( 10,-0.5){\line(0,1){0.5}}
   \put(  8,-2.5){\makebox(4,2)[c]{\large    0.1\AA }}
   \put( 10.56,0.0){\line(0,1){1.0}}
   \put(  9.56,1.0){\makebox(2,2)[c]{\Large\sf K }}
   \put( 10.80,0.0){\line(0,1){1.0}}
   \put( 10.80,3.0){\line(0,1){1.0}}
   \put(  9.80,4.0){\makebox(2,2)[c]{\Large\sf ~O$_3$ }}
   \put(  5.67,0.0){\line(0,1){4.0}}
   \put(  4.67,4.0){\makebox(2,2)[c]{\Large\sf ~O$_{1,2}$ }}
   \put( -8.26,0.0){\line(0,1){3.0}}
   \put( -9.26,3.0){\makebox(2,2)[c]{\Large\sf Nb }}
   \end{picture}
 }
 \multiput(15,4)(0,1){30}{\line(0,1){0.5}} 
 \put(10, 1){\makebox(10,2)[c]{\sf center of mass}}
 \end{picture}
}
\end{picture}
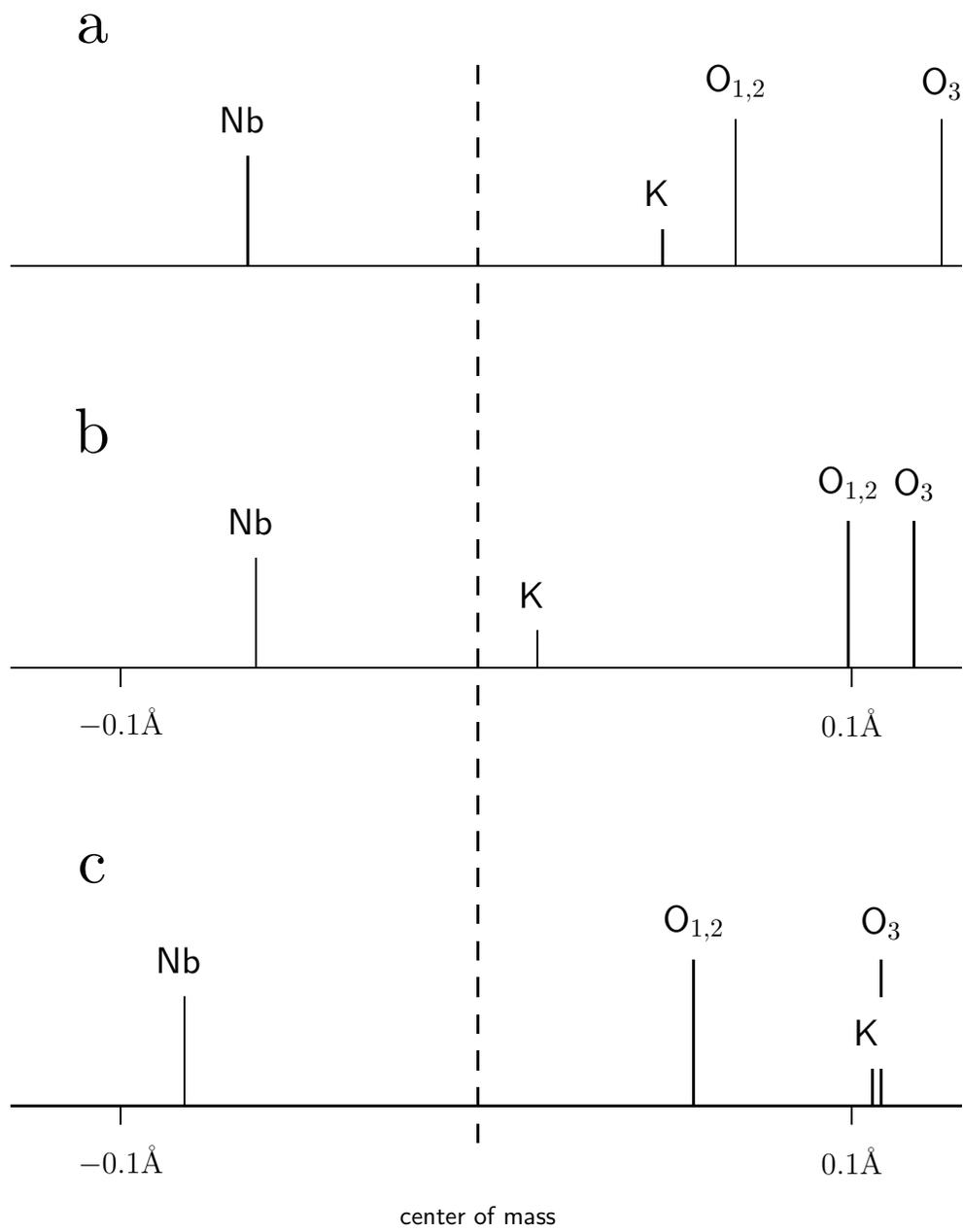
\end{document}